\documentclass[11pt, oneside]{article}
\usepackage[utf8]{inputenc}
\usepackage{amsmath}
\usepackage{parselines}
\usepackage{geometry}   
\usepackage[pdftex]{graphicx}			
\usepackage[usestackEOL]{stackengine}
\usepackage{wrapfig}
\usepackage{float}

\usepackage{setspace}
\usepackage{longtable}
\usepackage{verbatim}
\usepackage{lipsum}

\title{Bayesian Analysis of Stochastic Volatility Model using Finite Gaussian Mixtures with Unknown Number of Components}
\author{Soham Mukherjee\\
\small \textit{School of Mathematics and Statistics, University of Hyderabad}} \normalsize

\graphicspath{{Images/}}
\begin{document}

\maketitle

\begin{abstract}
Financial studies require volatility based models which provides useful insights on risks related to investments. Stochastic volatility models are one of the most popular approaches to model volatility in such studies. The asset returns under study may come in multiple clusters which are not captured well assuming standard distributions. Mixture distributions are more appropriate in such situations. In this work, an algorithm is demonstrated which is capable of studying finite mixtures but with unknown number of components. This algorithm uses a Birth-Death process to adjust the number of components in the mixture distribution and the weights are assigned accordingly. This mixture distribution specification is then used for asset returns and a semi-parametric stochastic volatility model is fitted in a Bayesian framework. A specific case of Gaussian mixtures is studied. Using appropriate prior specification, Gibbs sampling method is used to generate posterior chains and assess model convergence. A case study of stock return data for State Bank of India is used to illustrate the methodology.
\end{abstract}
\textit{MSC2020-Mathematics Subject Classifications} 62F15

\makeatletter{\renewcommand*{\@makefnmark}{}
\footnotetext{I wish to express my sincere gratitude to Prof. Diganta Mukherjee for providing the NSE-Nifty data on State Bank of India, insightful reviews and continued support.}}

\section{Introduction}
The analysis of stock market has always been a topic of constant interest in the world of financial researchers. Assessment of risk before investing in a stock is a necessity. The most popular way involving statistical approach is using the time series data of stock returns and model the volatility. Using such models, the underlying volatility are studied to assess the risk involved in a particular stock to make an informed decision. One of the popular models involve the evolution of volatility deterministically through the GARCH class of models by Engle (1982) and Bollerslev (1987). An alternative approach involve modelling the volatility probabilistically through a state-space model where the logarithm of the squared volatility or the latent states, follow an AR (1) process. This specification is known as Stochastic Volatility model first developed by Taylor (1982). Further works on similar model specifications include Hull and White (1987), Chesney and Scott (1989), Taylor (1986, 1994), Jacquier et al. (1994) and Shepherd (1996). Parametric and semi-parametric estimation methods of SV models include works of  Harvey et al. (1994), Carter and Kohn (1994), Kim et al. (1998), and Omori et al. (2007). Except the first, all three specify and demonstrate SV models using a Gaussian mixture approximation. To implement mixture models, the number of components need to be specified. Use of infinite component DPMs are popular since they represent the most general case of mixtures but comes at a cost of enhanced analytical complexity. A good balance is to assume that the mixture components are finite but unknown. The most important works like Richardson and Green (1997), Stephens (2000a) and Frühwirth-Schnatter (2006) demonstrate how to adjust and optimize the number of components for model specifications. Majority of these works involve some sort of Bayesian computation and simulation. Kim, Shepherd and Chib (1998) briefly mentioned how semiparametric SV models could be implemented using Reversible-Jump MCMC developed by Richardson and Green (1997). Later, Asai (2009) demonstrated use of mixture SV models for return volatility modeling by fixing a 2 component mixture. The Markov Chain Monte Carlo methods of inferences used were discussed in Gelfand and Smith (1990), Chib and Greenberg (1996) and Gilks, Richardson and Spiegelhalter (1996).\\
\newline
The case study data for NSE-Nifty State Bank of India stocks had observations coming from clusters of Gaussian population which is represented using finite Gaussian mixtures in this work. The data posed some typical challenges that are usually related to Bayesian inference for mixture distribution. Along with significant computational resource and time, there were issues of label switching for posterior means. Label switching is common for Bayesian mixture analysis where the swapping of component parameters due to permutations while simulation result in multiple posterior maxima. Since, the standard Bayesian method is to estimate parameters from marginal posterior means, swapping of mixture component parameter causes serious identification issues. Richardson and Green (1997) used ordering of posterior means as a constraint to deal with label switching. In Stephens (2000b), it is shown that these constraints in general fails to solve the problem and relabelling algorithms need to be implemented. Further, according to Diebolt and Robert (1994), ``Improper priors in mixture models result in improper posteriors." This would mean a divergent posterior chain and unusable estimates. So proper informative priors specified by previous works had to be used. Although using finite mixture distributions of stock returns and modelling their volatility is a popular approach, the method of dealing with finite unknown components and their optimization is usually done using RJMCMC method of Richardson and Green (1997). In this work, the alternate approach Birth-death MCMC developed by Stephens (2000a) is used which is both easy to implement and label switching invariant by construction. Further, a modified 3 dimensional point process for the Gaussian mixtures is used with varying mean, variance and the mixture weights. The algorithm developed by Stephens (2000a) is used with few adjustments in order to incorporate the Bayesian SV model via Gibbs sampling steps and priors used by Kim, Shepherd and Chib (1998). This approach brings a new outlook in the field of Bayesian semi-parametric SV model which is illustrated.\\
\newline
In this paper, the SV model as well as its finite Gaussian mixture specification is demonstrated in the section 2.1. In section 2.2, the algorithm along with the adjustments for finite Gaussian mixtures with unknown number of components are determined and discussed in detail. The prior specifications and the Gibbs sampling steps for the Bayesian SV model is demonstrated in section 2.3. Using these, the case study is used for illustration purpose in section 3 to demonstrate model fitting and performance followed by a discussion in section 4.

\section{Methodology}
The methodology section is divided into three parts. Section 2.1 describes the SV model that is used in this paper alongside the finite mixture specifications. 2.2 demonstrates the setup and algorithm to determine the number of components of the mixture distribution which is achieved by using a Birth-Death MCMC process. Finally, 2.3 explores the implementation of the determined number of components and its usage to obtain the posterior of the required parameters to fit a stochastic volatility model.
\subsection{Model}
The simple or canonical stochastic volatility model is considered for this work.
\begin{equation}
    y_t=\epsilon_t\exp(h_t/2)  
\end{equation}
\begin{equation}
     h_t=c+\phi(h_{t-1}-c)+\eta_t
\end{equation}
where $|\phi|<1$ and $y_t$ is asset return at $t$. $exp(h_t/2)$ is volatility, so $h_t$ is log of squared volatility. $\epsilon_t\sim iid$ with $E(\epsilon_t)=\mu_{\epsilon_t}$ and $Var(\epsilon_t)={\sigma_{\epsilon_t}}^2$, $\eta_t \sim iid N(0,\sigma_\eta^2)$. $\epsilon_t$ and $\eta_s$ are assumed to be mutually independent $\forall t,s$. $\epsilon_t$ is used to model $y_t$. This form of the model has been used by Taylor (1986, 1994), Hull and White (1987), Chesney and Scott (1989), Shepherd (1996), Ghysels, Harvey and Renault (1986), Jacquier, Polson and Rossi (1994), Kim, Shephard and Chib (1998). These above cited works discusses its basic econometric properties as well as estimation procedures of SV models. The same form is also used by Asai (2009) where author discusses its extension using two-component gaussian mixtures.
In this work, $\epsilon_t$ in the canonical model is assumed to follow mixture distribution with $k$ components (where $k$ is unknown). It is denoted by the usual form of a mixture distribution
\begin{equation}
 p(\epsilon_t|\pi,\mu,\sigma)=\pi_1f(\epsilon_t;\mu_1,\sigma_1)+...+\pi_kf(\epsilon_t;\mu_k,\sigma_k)  
\end{equation}
Literature on this form of mixtures have been discussed in the works of Frühwirth-Schnatter (2006). It is further assumed that $\epsilon_t$ is a finite mixture of univariate normals based on the discussions provided by Fama (1965) with adjustments made in the form of varying means in this case. So, $f(\epsilon_t;\mu_i,\sigma_i)$ is $N(\mu_i,\sigma_i^2)$ in the above notation.

\subsection{Determining the number of components}
To address the unknown components and weights, the Birth-Death MCMC by Stephens (2000a) is followed. Using the notations of a point process, $\epsilon_t$ is written in terms of $(\pi_i,\mu_i,\sigma_i)$ in the form $\epsilon_t=\{(\pi_1,\mu_1,\sigma_1),...,(\pi_k,\mu_k,\sigma_k)\}$ which belongs to parameter space $\Omega_k$. The author denotes the Births and Deaths as follows. When a birth occurs at $(\pi,\mu,\sigma)$, then the process jumps to
\begin{equation}
    \epsilon_t\cup(\pi,\mu,\sigma)=\{(\pi_1(1-\pi),\mu_1,\sigma_1),...,(\pi_k(1-\pi),\mu_k,\sigma_k),(\pi,\mu,\sigma)\} \in \Omega_{k+1}
\end{equation}
In case of a death at $(\pi_i,\mu_i,\sigma_i)\in\epsilon_t$, the process jumps to
\begin{multline}
     \epsilon_t\setminus(\pi_i,\mu_i,\sigma_i)=\{(\pi_1/(1-\pi_i),\mu_1,\sigma_1),...,(\pi_{i-1}/(1-\pi_i),\mu_{i-1},\sigma_{i-1}),\\ (\pi_{i+1}/(1-\pi_i),\mu_{i+1},\sigma_{i+1}),...,(\pi_k/(1-\pi_i),\mu_k,\sigma_k)\} \in \Omega_{k-1}  
\end{multline}
A birth increases number of components by one and a death decreases number of components by one. The entire setup is defined in a way that births and deaths are inverse mechanisms and the weights sum up to unity. When the process is at $\epsilon_t\in\Omega_k$, births and deaths occur as independent Poisson process. Births occur at overall rate $\beta(\epsilon_t)$ which is chosen with density $b(\epsilon_t;(\pi,\mu,\sigma))$. On the other hand, a point dies independently of others as a Poisson process with rate $\delta_j(\epsilon_t)=d(\epsilon_t\setminus(\pi_j,\mu_j,\sigma_j);(\pi_j,\mu_j,\sigma_j))$. The overall death rate is $\delta(\epsilon_t)=\sum\delta_j(\epsilon_t)$. Using the specific hierarchical prior on the parameters $(k,\pi,\mu,\sigma)$ used in Stephens (2000a) along with a density $r(k,\pi,\mu,\sigma)$ the mixture specification becomes invariant under relabeling of components which will eventually result in a more practical posterior. It is worthwhile to note that due to this specification, no additional conditioning on the parameters, for example, ordering of means which was used in Richardson and Green (1997) is unnecessary. Assuming $\pi$ and $\sigma$ are a priori independent and identically distributed from a distribution with density $\tilde{p}(\mu,\sigma)$, then
\begin{equation}
    r(k,\pi,\mu,\sigma)=p(k)\tilde{p}(\mu_1,\sigma_1)...\tilde{p}(\mu_k,\sigma_k)
\end{equation}
A special case of Diebolt and Robert (1994), Richardson and Green (1997) and Stephens (2000a) is to use mixtures of univariate normals. The likelihood is specified as follows. 
\begin{equation}
    L(k,\pi,\mu,\sigma)=p(y_t|k,\pi,\mu,\sigma)=\prod_{j=1}^n[\pi_1f(\epsilon_{tj};\mu_1,\sigma_1)+...+\pi_kf(\epsilon_{tj};\mu_k,\sigma_k)]
\end{equation}
As per the construct, this is invariant under permutations of component labels. In case of financial studies with asset returns, $y_t$ is the asset return data. Using the theorem proposed by Stephens (2000a), the posterior 
\begin{equation}
    p(k,\pi,\mu,\sigma| y_t)\propto L(k,\pi,\mu,\sigma)r(k,\pi,\mu,\sigma)
\end{equation}
has stationary distribution provided $b$ and $d$ satisfy 
\begin{equation}
    (k+1)d(\epsilon_t;(\pi,\mu,\sigma))r(\epsilon_t\cup(\pi,\mu,\sigma))L(\epsilon_t\cup(\pi,\mu,\sigma))k(1-\pi)^{k-1}
    =\beta(\epsilon_t)b(\epsilon_t;(\pi,\mu,\sigma))r(\epsilon_t)L(\epsilon_t)
\end{equation}
As with any other Bayesian implementation, prior specification is one of the most important aspect of a methodology. In this paper, the priors for the special case by Richardson and Green (1997) and Stephens (2000a) is used. 
A truncated Poisson prior is assumed for number of components $k$.
\begin{equation}
    p(k)\propto \frac{\lambda^k}{K!};(k=1(1)10, say)
\end{equation}
where $\lambda$ is a constant. For the rest of the parameters, the following prior specifications are considered.
\begin{equation}
    \beta \sim \Gamma(2l,(2m)^{-1})
\end{equation}
\begin{equation}
    \pi \sim \mathcal{D}(\gamma)
\end{equation}
\begin{equation}
    \mu_i \sim N(\zeta,\tau^{-1})
\end{equation}
\begin{equation}
    \sigma_i^{-1}|\beta \sim \Gamma(2\alpha,(2\beta)^{-1})
\end{equation}
where $\beta$ is a hyperparameter; $\mathcal{D}(\mu)$ denotes the symmetric Dirichlet distribution with density 
\begin{equation}
    \frac{\Gamma(k\gamma)}{\Gamma(\gamma)^k} \pi_1^{\gamma-1}...\pi_{k-1}^{\gamma-1}(1-\pi_1-...-\pi_{k-1})^{\gamma-1}
\end{equation}
$\zeta$ is the midpoint of the observed interval of variation in the data, $R$ is the length of the interval, $\tau=\frac{1}{R^2}$, $\alpha=2$, $l=0.2$, $m=\frac{100l}{\alpha R^2}$, $\gamma=1$. Using these, a Markov chain with suitable stationary distribution is simulated.
The generalized algorithm is implemented starting with the initial specification of $\epsilon_t=\{(\pi_1,\mu_1,\sigma_1),...,(\pi_k,\mu_k,\sigma_k)\}\in \Omega_k$. Using the priors, the following steps are performed.
\begin{enumerate}
    \item {The birth rate $\beta(\epsilon_t)=\lambda_b$ is specified by the researcher}
    \item{Using the birth rate and (9) the death rate is calculated for each component as 
    \begin{equation}
        \delta_j(\epsilon_t)=\lambda_b{\frac{(L(\epsilon_t\setminus(\pi_j,\mu_j,\sigma_j))}{L(\epsilon_t)} }{\frac{p(k-1)}{kp(k)} }
    \end{equation}}
    \item{The total death rate is calculated as $\delta(\epsilon_t)=\sum_j \delta_j(\epsilon_t)$}
    \item{The time to next jump is simulated from exponential distribution with mean $1/(\beta(\epsilon_t)+\delta(\epsilon_t))$}
    \item{The birth and death probabilities are computed as $\frac{\beta(\epsilon_t)}{\beta(\epsilon_t)+\delta(\epsilon_t)}$ and $\frac{\delta(\epsilon_t)}{\beta(\epsilon_t)+\delta(\epsilon_t)}$ respectively.}
    \item{$\epsilon_t$ is adjusted according to birth or death as defined in (5) and (6). Birth for the point $(\pi,\mu,\sigma)$ is obtained using $b(y;(\pi,\mu,\sigma))=k(1-\pi)^{k-1}\tilde{p}(\mu,\sigma)$ by simulating $\pi$ from $k(1-\pi)^{k-1}$ and $(\mu,\sigma)$ from $\tilde{p}(\mu,\sigma)$ independently. Death for a component is obtained from $(\pi_i,\mu_i,\sigma_i)\in \epsilon_t$ selected with probability $\delta_j(\epsilon_t)/\delta(\epsilon_t)$}
    \item{Steps 2 to 6 are repeated until convergence or max iterations}
\end{enumerate}
Alternatively, the Step (7) could be skipped, that is the above steps are performed only once in order to obtain an updated initial value of $(k, \pi, \mu, \sigma)$ denoted by $(k^{(j)},\pi^{(j)},\mu^{(j)},\sigma^{(j)})$. Using Gibbs sampler,
\begin{enumerate}
    \item{Sample $\mu^{(j+1)}$ from $p(\mu|k^{(j+1)},\pi^{(j)},\sigma^{(j)},y_t)$}
    \item{Sample $\sigma^{(j+1)}$ from $p(\sigma|k^{(j+1)},\mu^{(j+1)},\pi^{(j)},y_t)$}
    \item{Sample $\pi^{(j+1)}$ from $p(\pi|k^{(j+1)},\mu^{(j+1)},\sigma^{(j+1)},y_t)$}
\end{enumerate}
Repeating this until model convergence would provide the required posterior sample of $(k, \pi, \mu, \sigma)$ which would then be presented alongside the posteriors of SV model parameters demonstrated in the following section 2.3.
\subsection{Fitting the stochastic volatility model}
Having simulated the number of components and the weights, all that is left is simulating the latent volatility model. Following the works of Kim et al. (1998) and Asai (2009), the priors for $(c,\phi,\sigma_\eta^2)$ follows $c \sim N(0,10)$, $(\phi+1)/2 \sim Beta(\phi_2,\phi_1),\sigma_\eta^2 \sim IG(\sigma_r/2,S_\sigma/2)$ where $IG$ denotes inverse gamma distribution; $\phi_1=20$, $\phi_2=1.5$, $\sigma_r=5$, $S_\sigma=0.01\sigma_r$. The prior distribution of $h_0$ is set to be the unconditional distribution of $h_t$ that is $h_0 \sim N(c,\sigma_\eta^2/(1-\phi^2))$. The Gibbs sampler is used to generate samples using the parameters $(c,\phi,\sigma_\eta^2)$ and log-squared volatility $h_t$ given the data $y_t$. The following steps are followed.
\begin{enumerate}
    \item {Initialize $h$ and $(c,\phi,\sigma_\eta^2)$}
    \item{Sample $h_t$ from $h_t|h_{\setminus t},c,\phi,\sigma_\eta^2,y_t,t=1(1)n$}
    \item{Sample $\sigma_\eta^2|y,h_t,\phi,c$}
    \item{Sample $\phi|h_t,c,\sigma_\eta^2$}
    \item{Sample $c|h_t,\phi,\sigma_\eta^2$}
    \item{Re-iterate step 2}
\end{enumerate}
The Gibbs sampler will be required to repeat steps through 2 to 5 many thousand times or even higher to generate samples from $c,\phi,\sigma_\eta^2,h_t|y_t$.
This should provide an appropriate posterior sample for the Stochastic Volatility Mixture Gaussian model.

\section{Illustration}
The algorithm is implemented in the case study of NSE-Nifty covering 6 working days of September 2017 stocks of State Bank of India which contains about 409 observations in 10 minute time intervals. The various exploratory plots for the return data are \newpage

\begin{figure}
    \centering
    \includegraphics[scale=0.5]{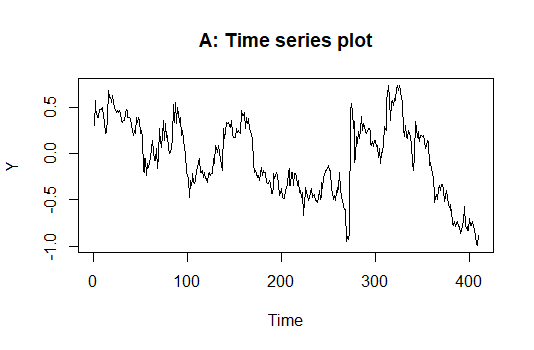}
    \includegraphics[scale=0.5]{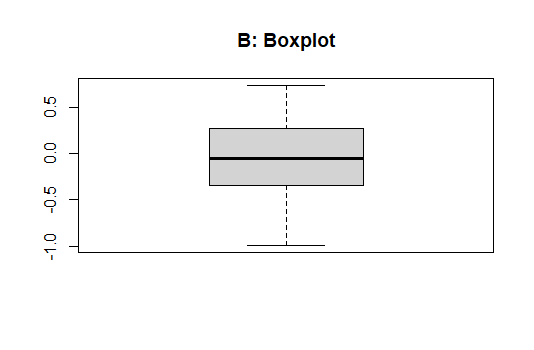}\\
    \includegraphics[scale=0.5]{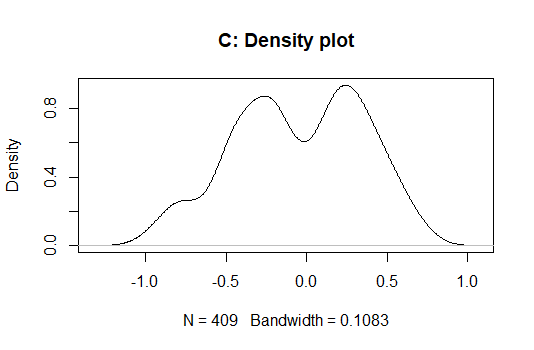}
    \includegraphics[scale=0.5]{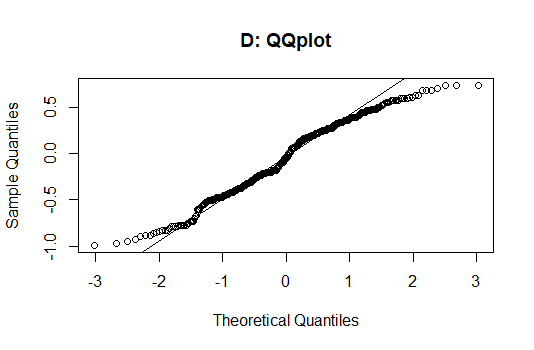}\\
    \caption{Summary Plots}
    \label{Fig1}
\end{figure}
The plots show the range of return values for the case study. Additionally the data is free of outliers. The density plots provide indication of presence of mixtures and the QQplot indicates that the returns are leptokurtic. The leptokurtic nature of asset returns in financial studies are recognized since Mandelbrot (1963) and Fama (1965). Fama (1965) further specifies a possible explanation of heavy tailed nature could be a mixture of normal distributions with same mean and different variances. The density plot here however shows distinct modal peaks. So, it will be adequate to consider different means as well.\\\\
Performing the algorithm in 2.2 once to obtain a set of initial values, a Gibbs sampling is implemented to obtain a posterior sample of number of components $k$. The simulation is run for 1000 iterations with 100 burn-in. An easy-to-use R package \textit{bmixture} by Mohammadi (2021) will provide similar and necessary implementation for the initial step. The number of components are determined to be $$(2,1,2,1,...,8,9,8,7,6)$$
The number of components is considered $k=6$ and the initial values of the mixture components generated by the Birth-Death MCMC algorithm is given as

\begin{align*}
\pi&=(0.0604687,0.1310388,0.1116203,0.3106168,0.2236884,0.1625669)\\
\mu&=(-0.8032387,-0.4422737,-0.2452619,-0.1486318,0.2454454,0.4692405)\\
\sigma&=(0.004225172,0.002882527,0.006002868,0.101062441,0.011680629,0.018423264)    
\end{align*} 
Using the Gibbs sampling steps in 2.3, the SV model is fitted using JAGS in R. The posterior values of the parameters are obtained and recorded when a convergent model is observed. The Gelman-Rubin statistic developed by Gelman and Rubin (1992) is used to assess model convergence. The maximum value for Gelman-Rubin statistic for the obtained model is 1.063 which is acceptable for a Bayesian model assessment as a numerical convergence indication. The posterior means are recorded which can be used to specify the SV model. A comparison with a non-mixture specified Bayesian SV model with similarly tuned Gibbs sampler as well as same prior specifications is shown.
\begin{table}[htb]
    \centering
    \caption{SV and Mixture SV comparison}
    \begin{tabular}{|*{5}{c|}} \hline
    Parameters & \multicolumn{2}{c|}{SV} 
         & \multicolumn{2}{c|}{Mixture SV} \\
         \hline
          & Mean & Rhat & Mean & Rhat \\\hline
         $c$ & -0.343  & 1.193 & -0.136  & 1.028  \\
         $\phi$ & 0.949 & 1.128 & 0.983  &  1.051\\
         $\sigma_{\eta_t}$ & 0.007 & 1.022 &  0.215  & 1.063 \\
         deviance & 1143.911  & 1.060   & -798.982  & 1.009    \\
         \hline
    \end{tabular}
    \label{Table1}
\end{table}

This table shows comparison between the posterior means of SV parameters for the two models. Without the mixture specification that is by considering the data is not comprised of a cluster of observations, the numerical convergence of the model is not achieved which is indicated by the Gelman-Rubin statistic value of $>1.1$ for some parameters. Even the deviance posterior mean value is much lower for the Mixture SV model. This provides the insight that mixture specification was indeed necessary. The complete summary table along with diagnostics for Mixture SV model is provided in the Appendix (Table 2) of this paper. The posterior means of mixture component parameters are observed. Specifically, it is interesting to see the variance of each of the components is very close. This indicates that the components are similarly dispersed. In the summary table, the inverse of variance is provided and its small values indicate somewhat large variance. The means of the components are very close to zero which captures and represents the asset return data successfully. The weights assigned to the clusters add up to unity satisfying the mixture specification conditions. The weight assigned to the first component which is $N(-0.008, 17.2413)$ is the largest among the six components. The fitted SV model is given as in (1) and (2) with $c=-0.136$, $\phi=0.983$ and $\eta_t \sim iid N(0,0.215)$.
For visual proof of convergence, trace plots are obtained for the posterior samples of model parameters. The trace plots are provided in the Appendix (Figure 2). Couple of parameters show less amount of mixing of two posterior chains compared to others. A natural solution is to simulate model for more iterations. That would provide more posterior samples, which will in turn show better quality trace plots with converged chains and better mixing but it would also take much longer time as well has more computation power and resources. Since, numerical convergence is achieved, more simulations are not performed. The density plots provided in the Appendix (Figure 3) for the posterior sample show some deviation from the prior specifications. The visual difference is not very large which means the prior specifications are appropriate. The slight deviations also speak about the quality of the data since the data had enough information to influence the prior choices.
\section{Discussion}
This paper demonstrates a Bayesian semi-parametric SV model for situations where target variable comes from a mixture distribution with finite but unknown number of components. Fixing the number of components without any hard evidence does not really reflect a realistic scenario. The usage of finite mixtures by introducing a flexibility in number of components provides a practical approach although the computation becomes comparatively resource intensive. In this work the Birth-Death MCMC is used instead of the more popular RJMCMC due to the ease of implementation and intuitiveness. Further, the algorithm being label switching invariant makes the posterior means more reliable without having to introduce additional parameter constraints. The number of components essentially comes from the appropriate choice of priors as well as a user specified birth rate. This setup results in death rate greater than birth rate most of the time. This means that the number of components will die faster than their birth. To compensate this, an appropriate starting value for the number of components is specified for the case study using the crude estimate of distinct modes in the density graph of the stock returns. The algorithm demonstrated holds for more general distributions (non-Gaussian) with more parameters. The point process specification will need to be adjusted to reflect a higher dimension and the algorithms can be modified accordingly along with appropriate priors. The mixture components posterior estimates show Gaussian means close to zero along with very close variance. The dispersion of the clusters of asset returns can be claimed similar. The SV model obtained in the case study shows model convergence and the fitted model parameters satisfy all model constraints like sum of mixture weights to unity and $\phi<1$. These results make the fitted SV model for the case study a practical and usable model. The sample size of 409 is used due to computational resource limitations, however satisfactory results are achieved. Multiple comparisons could not be shown since firstly, due to novelty of the work, an exact competitive model was difficult to find in existing literature. The non-mixture specified model is compared and the limitations are highlighted in the previous section. Models from other Bayesian samplers are not compared due to the compatibility issues for the specific Bayesian packages in R. Due to this, such complex models from different samplers are very difficult to compare and so it is kept outside the scope of this paper. Potential extensions could include a more robust mixture component optimization which would be less sensitive to starting point of $k$. This may be explored by modifying user specified birth rates and replacing it with data dependent birth rates. Although the algorithm has provisions for non-Gaussian distributions, only the specific case of univariate mixture normals is studied. Other potential distributions may be explored but that would also require different prior specification through rigorous prior elicitation methods which are beyond the scope of this paper.
\section*{References}
\singlespacing
\begin{enumerate}
    \item{Asai. M (2009) Bayesian Analysis of Stochastic Volatility Models with Mixture-of-Normal Distribution, \textit{Mathematics and Computers in Simulation}, \textbf{79(8)}, 2579-2596.}
    \item {Bauwens L., Hafner C. and Laurent S. (2012) Handbook of Volatility Models and their Applications, \textit{Wiley}.}
    \item{Bollerslev, T. (1987) A conditional heteroskedastic time series model for speculative prices and rates of return, \textit{Review of Economics and Statistics}, \textbf{69}, 542-547.}
    \item{Carter C. K. and Kohn R. (1994) On Gibbs sampling for state space models, \textit{Biometrika}, \textbf{81}, 541-553.}
    \item{Chesney M. and Scott L. O. (1989) Pricing European options: a comparison of the Black Scholes model and a random variance model, \textit{Journal of Financial and Qualitative Analysis}, \textbf{24}, 267-284.}
    \item{Chib S. and Greenberg E. (1996) Markov chain Monte Carlo simulation methods in econometrics, \textit{Econometric Theory}, \textbf{12}, 329-335.}
    \item{Diebolt J. and Robert C. (1994) Estimation of Finite Mixture Distribution through Bayesian Sampling, \textit{Journal of Royal Statistical Society B}, \textbf{56(2)}, 363-375.}
    \item{Engle, R. (1982). Autoregressive Conditional Heteroscedasticity with Estimates of the Variance of United Kingdom Inflation, \textit{Econometrica}, \textbf{50}, 987-1007.}
    \item{Fama E. (1965) The Behaviour of Stock-Market Prices, \textit{The Journal of Business}, \textbf{38}, 34–105.}
    \item{Fruhwirth-Schnatter S. (2006) Finite Mixture and Markov Switching Models, \textit{Springer}}
    \item{Gelfand A. and Smith A. F. M. (1990) Sampling-based approaches to calculating marginal densities,\textit{Journal of the American Statistical Association}, \textbf{85}, 398-409.}
    \item{Gelman. A. and Rubin. D. (1992). Inference from Iterative Simulation Using Multiple Sequences.\textit{Statistical Science}, \textbf{7(4)}, 457-472.}
    \item{Ghysels E., Harvey A. C. and Renault E. (1996) Stochastic volatility, in C. R. Rao and G. S. Maddala (eds.) \textit{Statistical Methods in Finance}, 119-191.}
    \item{Gilks W. K., Richardson S. and Spiegelhalter D. J. (1996) Markov Chain Monte Carlo in Practice, \textit{Chapman and Hall}.}
    \item{Harvey A. C, Ruiz E. and Shephard N. (1994) Multivariate stochastic variance models, \textit{Review of Economic Studies}, \textbf{61}, 247-264.}
    \item{Hull, J. and A. White (1987) The pricing of options on assets with stochastic volatilities, \textit{Journal of Finance}, \textbf{42}, 281–300.}
    \item{Jacquier E., Polson N. G. and Rossi P. E. (1994) Bayesian analysis of stochastic volatility models (with discussion), \textit{Journal of Business and Economic Statistics}, \textbf{12}, 371-417.}
    \item{Kim S., Shephard N. and Chib S. (1998) Stochastic Volatility: Likelihood Inference and Comparison with ARCH Models, \textit{Review of Economic Studies}, \textbf{65}, 361-393.}
    \item{Mandelbrot B. (1963) The Variation of Certain Speculative Prices, \textit{The Journal of Business}, \textbf{36}, 394–419.}
    \item{Mohammadi R (2021) bmixture: Bayesian Estimation for Finite Mixture of Distributions, \textit{R package}, version \textbf{1.7}.}
    \item{Omori Y., Chib S., Shephard N., and Nakajima J. (2007). Stochastic volatility with leverage: Fast and efficient likelihood inference, \textit{Journal of Econometrics}, \textbf{140}, 425–449.}
    \item{Richardson S. and Green P. (1997) On Bayesian analysis of Mixtures with an Unknown Number of Components, \textit{Journal of Royal Statistical Society B}, \textbf{59(4)}, 731-792.}
    \item{Shephard N. (1996) Statistical aspects of ARCH and stochastic volatility, D. R. Cox, O. E. Barndoff-Nielson and D. V. Hinkley (eds.), Time Series Models in Econometrics, Finance and Other Fields, \textit{Chapman and Hall}, 1-67.}
    \item{Stephens M. (2000a) Bayesian Analysis of Mixture Models with an Unknown Number of Components-An Alternative to Reversible Jump Methods, \textit{The Annals of Statistics}, \textbf{28(1)}, 40-74.}
    \item{Stephens M. (2000b) Dealing with Label Switching in Mixture Models, \textit{Journal of the Royal Statistical Society B}, \textbf{62(4)}, 795–809.}
    \item{Taylor S. J. (1982) Financial returns modelled by the product of two stochastic processes — a study of daily sugar prices 1961-79, In O. D. Anderson (Ed.), \textit{Time Series Analysis: Theory and Practice}, 1, pp. 203–226, Amsterdam: North Holland.}
    \item{Taylor S. J. (1986) Modelling Financial Time Series, \textit{Wiley}.}
    \item{Taylor S. J. (1994) Modelling Financial Time Series, \textit{Mathematical Finance}, \textbf{4}, 183-204.}
\end{enumerate}
\setstretch{1.3}
\section*{Appendix}
\floatstyle{plaintop}
\restylefloat{table}
\begin{table}[htb]
    \centering
    \begin{tabular}{|l|l|l|l|l|l|l|l|l|}
\hline
Parameter            & mu.vect  & sd.vect & 2.50\%   & 25\%     & 50\%     & 75\%     & 97.50\%  & Rhat  \\\hline
$c$           & -0.136   & 0.308   & -0.736   & -0.352   & -0.145   & 0.07     & 0.485    & 1.028 \\
$\mu_1$   & -0.008   & 0.063   & -0.116   & -0.049   & -0.029   & 0.043    & 0.108    & 1.003 \\
$\mu_2$   & -0.001   & 0.066   & -0.119   & -0.046   & -0.023   & 0.052    & 0.115    & 1.001 \\
$\mu_3$   & -0.001   & 0.072   & -0.123   & -0.047   & -0.024   & 0.049    & 0.139    & 1.010 \\
$\mu_4$   & -0.009   & 0.064   & -0.115   & -0.049   & -0.028   & 0.042    & 0.112    & 1.001 \\
$\mu_5$   & -0.005   & 0.068   & -0.125   & -0.05    & -0.027   & 0.043    & 0.12     & 1.001 \\
$\mu_6$   & -0.001   & 0.068   & -0.132   & -0.048   & -0.018   & 0.05     & 0.12     & 1.005 \\
$\phi$         & 0.983    & 0.012   & 0.942    & 0.983    & 0.986    & 0.989    & 0.994    & 1.051 \\
$\pi_1$   & 0.201    & 0.294   & 0.001    & 0.011    & 0.034    & 0.29     & 0.912    & 1.004 \\
$\pi_2$   & 0.175    & 0.286   & 0.001    & 0.01     & 0.029    & 0.168    & 0.907    & 1.000 \\
$\pi_3$   & 0.144    & 0.244   & 0.001    & 0.01     & 0.025    & 0.142    & 0.846    & 1.033 \\
$\pi_4$   & 0.188    & 0.297   & 0.001    & 0.011    & 0.031    & 0.208    & 0.922    & 1.008 \\
$\pi_5$   & 0.16     & 0.26    & 0.001    & 0.012    & 0.031    & 0.187    & 0.907    & 1.006 \\
$\pi_6$   & 0.132    & 0.242   & 0.001    & 0.009    & 0.026    & 0.082    & 0.899    & 1.008 \\
$\sigma_{\eta_t}$           & 0.215    & 0.031   & 0.159    & 0.192    & 0.212    & 0.235    & 0.282    & 1.063 \\
$\sigma^{-1}_1$ & 0.058    & 0.056   & 0.01     & 0.023    & 0.041    & 0.073    & 0.216    & 1.005 \\
$\sigma^{-1}_2$ & 0.057    & 0.052   & 0.009    & 0.024    & 0.042    & 0.073    & 0.199    & 1.005 \\
$\sigma^{-1}_3$ & 0.058    & 0.055   & 0.008    & 0.024    & 0.041    & 0.071    & 0.217    & 1.012 \\
$\sigma^{-1}_4$ & 0.058    & 0.056   & 0.008    & 0.024    & 0.041    & 0.069    & 0.212    & 1.004 \\
$\sigma^{-1}_5$ & 0.057    & 0.058   & 0.008    & 0.024    & 0.04     & 0.071    & 0.195    & 1.011 \\
$\sigma^{-1}_6$ & 0.058    & 0.057   & 0.007    & 0.023    & 0.039    & 0.068    & 0.221    & 1.012 \\
deviance    & -798.982 & 64.238  & -928.981 & -839.455 & -799.588 & -758.839 & -670.416 & 1.009\\
\hline
\end{tabular}
    \caption{Summary Diagnostic Table}
    \label{Table2}
\end{table}

\begin{figure}
    \centering
    \includegraphics[scale=0.6, angle=90]{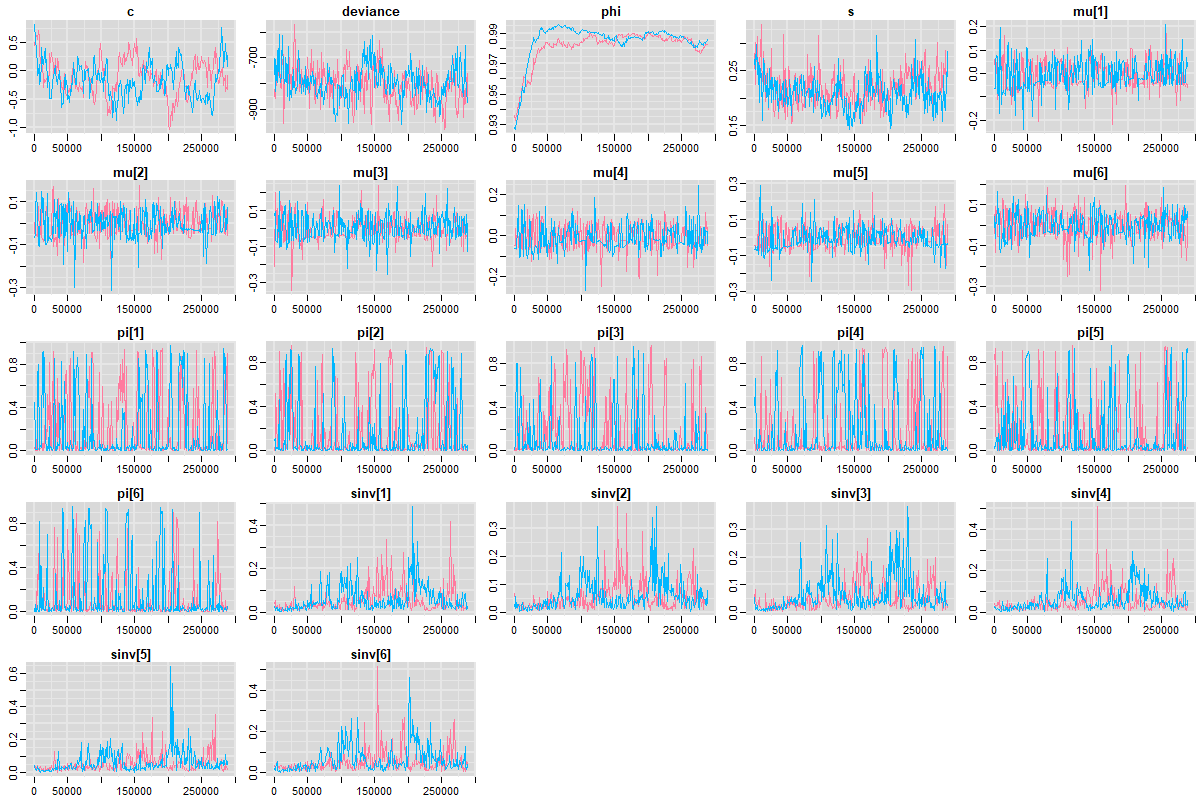}\\
    \caption{Trace Plots}
    \label{Fig2}
\end{figure}
    
\begin{figure}
    \centering
    \includegraphics[scale=0.6, angle=90]{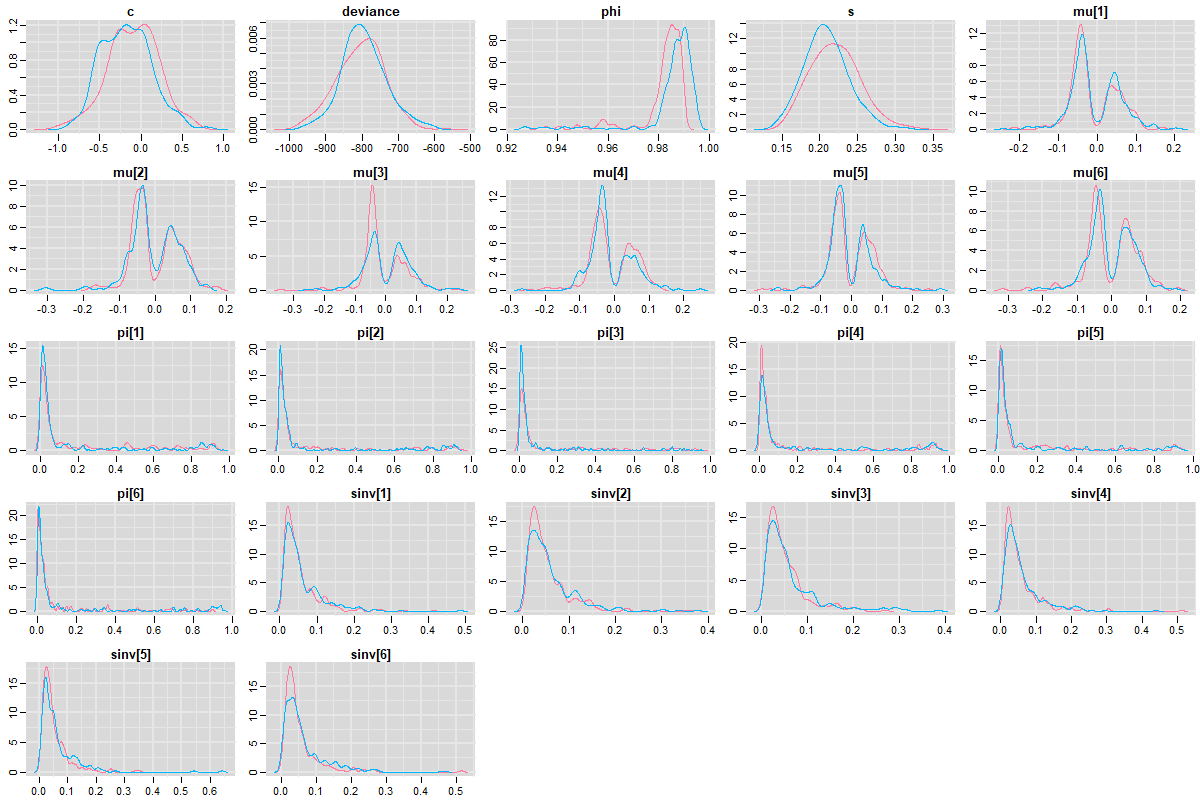}\\
    \caption{Posterior density plots}
    \label{Fig3}
\end{figure}
    
\end{document}